\documentclass[twocolumn,showpacs,amsmath,prl,superscriptaddress]{revtex4-1}
\usepackage{color}
\usepackage{graphicx}
\usepackage{epsfig,psfrag}
\usepackage{amsmath}
\usepackage{amssymb}
\usepackage{amsbsy}
\usepackage{amsthm}
\usepackage{amsfonts}
\usepackage{wasysym}
\usepackage{bbm}
\usepackage{tabularx}
\usepackage{euscript}
\usepackage{enumerate}
\usepackage{amsfonts}
\usepackage{exscale}
\usepackage{bbold}
\usepackage{float}
\usepackage{slashed}
\usepackage{subfigure}
\usepackage{comment}
\usepackage[colorlinks,citecolor=blue]{hyperref}

\newcommand{\bea}{\begin{eqnarray}}
\newcommand{\eea}{\end{eqnarray}}
\newcommand{\ba}{\begin{array}}
\newcommand{\ea}{\end{array}}

\newcommand{\nn}{\nonumber\\}

\newcommand{\Tr}{{\rm Tr}}

\def\bea{\begin{eqnarray}}
\def\eea{\end{eqnarray}}

\def\Tr{\mathrm{Tr}}

\def\nn{\nonumber}

\begin{document}

\title{Universal Prethermal Dynamics in Gross-Neveu-Yukawa Criticality}

\author{Shao-Kai Jian}
\affiliation{Institute for Advanced Study, Tsinghua University, Beijing 100084, China}
\affiliation{Condensed Matter Theory Center, Department of Physics, University of Maryland, College Park, Maryland 20742, USA}

\author{Shuai Yin}
\email{sysuyinshuai@gmail.com}
\affiliation{Institute for Advanced Study, Tsinghua University, Beijing 100084, China}
\affiliation{School of physics, Sun Yat-Sen University, Guangzhou 510275, China}

\author{Brian Swingle}
\email{bswingle@umd.edu}
\affiliation{Condensed Matter Theory Center, Maryland Center for Fundamental Physics, Joint Center for Quantum Information and Computer Science, and Department of Physics, University of Maryland, College Park MD 20742, USA}

\begin{abstract}
We study the prethermal dynamics of the Gross-Neveu-Yukawa quantum field theory, suddenly quenched in the vicinity of a critical point. We find that the universal prethermal dynamics is controlled by two fixed points depending on the size of the quench. Besides the usual equilibrium chiral Ising fixed point for a shallow quench, a dynamical chiral Ising fixed point is identified for a deep quench. Intriguingly, the latter is a non-thermal fixed point without any equilibrium counterpart due to the participation of gapless fermionic fields. We also find that in the scaling regime controlled by the equilibrium fixed point, the initial slip exponent is rendered negative if there are enough flavors of fermions, thus providing a unique signature of fermionic prethermal dynamics. We then explore the temporal crossover between the universal scaling regimes governed by the two universality classes. Possible experimental realizations are also discussed.
\end{abstract}
\date{\today}
\maketitle

{\it Introduction.---}Memory effects are ubiquitous phenomena in nature. The classic example is brain memory in living beings, but memory effects are also widespread in physics. In cosmology, the cosmic microwave background radiation can be regarded as a memory of the Big Bang. In condensed matter physics, memory effects often occur in relaxation dynamics~\cite{Calabrese2005,tauberbook04}. For example, in classical critical systems, the short-time critical dynamics remembers the initial state and affects the critical relaxation process during a macroscopic initial stage~\cite{janssen1988}. In this stage, the evolution of the system is called the critical initial slip and is characterized by an additional critical exponent, the initial slip exponent~\cite{janssen1988}. These short-time critical dynamics have been widely exploited in determining the critical point and critical exponents in classical systems~\cite{Li1995,Zheng1998}.

In the context of isolated quantum systems, a vibrant set of purely quantum memory phenomena have been studied. At long times, quantum chaotic systems are expected to effectively lose memory of their initial state, except for conserved quantities like the total energy. This is encoded in the celebrated eigenstate thermalization hypothesis which states that chaotic energy eigenstates look like equilibrium thermal states of the appropriate temperature~\cite{deutsch1991,srednicki1994,rigol2008,vengalattor2011,Dziarmaga2010}. However, some long-lived prethermal states, which bring in additional universal initial state information into the dynamics, have been discovered~\cite{wetterich2004,gring2012,langen2013,Eigen2018,mitra2018,Langen2016,ueda2018b}. Various causes for these effects have been proposed, including proximity to integrability~\cite{Kollar2011,Kastner2013,Marcuzzi2013,Smith2013,Bertini2015,Diehl2016}, existence of a dynamical phase transition after a global quench~\cite{Calabrese2006,Calabrese2007,Eckstein2009,Sciolla2010,Fabrizio2010,demler2011,Tsuji2013a,Tsuji2013b,Sciolla2013,Heyl2013,Sondhi2013,Silva2015}, emergence of a non-thermal fixed point~\cite{Berges2008,Gasenzer2011,Gasenzer2012,Berges2014,Berges2015,Erne2018,Oberthaler2018,Gasenzer2019,Fujimoto2019,Berges2019}, non-local initial entanglement effects~\cite{ueda2015,ueda2018}, and so on. Among these, short-time quantum critical dynamics have recently received considerable attention~\cite{Polkovnikov2013,Yin2014,Schmalian2014,Schmalian2015,mitra2015,Gambassi2015,mitra2016,marino2017}. As with its classical counterpart, a critical initial slip exponent can be defined to characterize the dynamical fingerprint of the initial state. However, unlike the classical case in which the short-time dynamics is completely controlled by a corresponding equilibrium critical point, short-time quantum critical dynamics in an isolated system can also be controlled by a dynamical fixed point, which is similar to a thermal fixed point for a purely bosonic field~\cite{mitra2015,Gambassi2015,mitra2016,marino2017}.

In the context of equilibrium criticality, gapless Dirac fermions, which appear in various systems including graphene, the surface of topological insulators, and Weyl semimetals, lead to exotic quantum phase transitions~\cite{sachdevbook}. The best known example is the chiral Ising universality class~\cite{nambu1961a,nambu1961b,gross1974,scherer2018}, in which the gapless Dirac fermion is coupled to a bosonic field via a Yukawa coupling. However, dynamical phase transitions in these systems have rarely been studied. Given ubiquity of Dirac systems in nature and the exotic universal physics associated to them, studies of the nonequilibrium behavior of these systems are urgently called for. Specifically, some important questions arise: To what extent is the prethermal dynamics affected by fermions? Can an associated dynamical fixed point always be cast as a thermal fixed point?

In this paper, we explore memory effects in Dirac systems. We study the prethermal dynamics of a Dirac system with $N$ flavors of two-component Dirac fermions~\cite{nambu1961a,nambu1961b,gross1974,scherer2018} after a sudden quench to the critical point. Using the Keldysh renormalization group (RG) analysis~\cite{kamenev2011}, we identify two non-trivial fixed points. The usual equilibrium chiral Ising fixed point (ECIFP) controls the near-equilibrium prethermal dynamics induced by a shallow quench, and a \textit{dynamical chiral Ising fixed point} (DCIFP), which is a \textit{non-thermal fixed point}, controls the prethermal dynamics after a deep quench. For both cases, we find that prethermal dynamics influenced by the initial condition is characterized by a critical initial slip exponent. Moreover, a negative initial slip exponent, induced by the fermionic degree of freedom, is found for shallow quenches. We also explore the temporal crossover between the two universal scaling regimes. Finally, the eventual long-time thermalization stage and possible experimental realizations are discussed.

\begin{table*}[t]
\centering
\caption{Critical initial slip exponents of the ECIFP and the DCIFP for the shallow quench and deep quench, respectively. Note the different definitions of $\epsilon$ in these two cases. $\eta$ and $\nu$ are also exhibited for comparison.}
  \begin{tabular}{c |cccc}
  \hline
  \hline
  ~~~~~~Prethermal universality class~~~~~~ & ~~~~~~~$\theta$~~~~~~~ &~~~$\eta (\equiv 2\eta_b)$~~~ & ~~~$\eta_f$~~~   & ~~~~~~~~$\nu^{-1}$~~~~~~~  \\
  \hline
 Equilibrium chiral Ising ($\epsilon\!=\!3\!-\!d$) & $\frac{3-37N+\sqrt{9+N(66+N)}}{24(3+N)} \epsilon$   & $\frac{N}{3+N} \epsilon$ & $\frac{1}{4(3+N)} \epsilon$  & $~~2\!-\! \frac{3+5N+\sqrt{9+N(66+N)}}{6+N}\epsilon$ \\
 Dynamical chiral Ising ($\epsilon\!=\!4\!-\!d$) & $\frac{\epsilon}{12}$  & $\mathcal O(\epsilon^2)$ & $\frac{\epsilon}{12}$  & $2 \!-\! \frac{\epsilon}3$ \\
  \hline
  \hline
  \end{tabular}
\label{exponent}
\end{table*}

{\it Quench protocol.---}The system consists of $N$ flavors of two-component Dirac fermions coupled to a real boson (which can be understood as the fluctuations of an order parameter) in $d$-spatial dimensions with Hamiltonian, 
\bea
	&& H(r,g,u) \!=\! \int_x \! \Big[ \frac12 \pi^2 + \frac12 (\nabla \phi)^2+ \frac{r}2 \phi^2 + \frac{u}{4!}\phi^4 \nn \\
		&& \quad\quad\quad\quad\quad + \sum_\alpha i \bar\psi_\alpha \gamma \cdot \nabla \psi_\alpha + g \phi \sum_\alpha \bar\psi_\alpha \psi_\alpha \Big] ,
\eea
where $\int_x \equiv \int d^d x$, $\phi$ and $\pi$ are the boson and its conjugate momentum, respectively. The fermion notation is as follows: $\psi_\alpha$ denotes Dirac fermions with flavor index $\alpha=1,...,N$, $\bar \psi \equiv \psi^\dag \gamma^0$, $\gamma \cdot \nabla \equiv \sum_{i=1}^d \gamma^i \partial_i$, and $\gamma^i$ are the gamma matrices. $r$ is the mass of the boson, which controls the distance to the critical point, while $g$, $u>0$ refer to the Yukawa coupling and four-boson coupling, respectively. For $t<0$, the initial state is the ground state of $H(\Omega_0^2, 0, 0) $. At $t=0$, the Hamiltonian is suddenly quenched to $H(r,g,u)$, and we are interested in the emergent universal behaviors after this global quench~\cite{supp}.

Without the coupling to fermions, $g=0$, it was shown that the nature of the universal prethermal dynamics depends on ratio of $\Omega_0$ and $\Lambda$, the UV energy cutoff: the system exhibits a dimensional crossover from an equilibrium ``quantum" Wilson-Fisher fixed point for a shallow quench $\Omega_0 \ll \Lambda$ to a dynamical ``classical" fixed point for a deep quench $\Omega_0 \gg \Lambda$~\cite{marino2017}. In lattice Dirac systems $\Lambda$ is set by the inverse lattice constant (in units where the Dirac velocity is one). It has also been shown that the ``classical" fixed point has some of the features of the equilibrium transition with an effective temperature determined by $\Omega_0$~\cite{mitra2015,Gambassi2015,mitra2016,marino2017}.

When the boson-fermion coupling is turned on, $g>0$, the Dirac fermions can affect the prethermal dynamics. For a quench of the type described above, the initial state information is contained in the free boson Keldysh propagator which reads~\cite{supp}
\bea
i D_K(k,t,t') \approx \frac{\omega_{0k}}{2\omega_k^2} [\cos \omega_k (t-t') \!-\! \cos \omega_k(t+t')],
\eea
where $\omega_{0k}^2 =k^2+\Omega_0^2$ and $\omega_k^2 = k^2 + r$, $k^2 \equiv \sum_{i=1}^d k_i^2$. Note that the second term explicitly breaks time translation invariance due to the quench. If $\Omega_0 \gg \Lambda$, $ i D_K\approx \frac{\Omega_0}{2\omega_k^2} [\cos \omega_k (t-t')- \cos \omega_k(t+t')]$, one can compare it with equilibrium Keldysh propagator at low temperature, $ i D_K \approx \frac{2T}{\omega_k^2}\cos \omega_k(t-t')$, which shows that $\Omega_0$ plays a role of effective temperature, $T_\text{eff} = \frac{\Omega_0}4$.

Because $\Omega_0$ sets an effective temperature scale, one may expect a ``quantum" to ``classical" crossover similar to the bosonic system~\cite{marino2017}. Moreover, given their inherently quantum nature, fermions are not expected to play a role in a classical phase transition~\cite{Stephanov,Hesselmann}. Thus, one might expect that the fermions effectively decouple with a vanishing Yukawa coupling for a deep quench $\Omega_0 \gg \Lambda$, but we will show that this is not the case. 

{\it RG analysis.---}Our analysis is based on a renormalization group (RG) calculation. In the Keldysh formalism~\cite{kamenev2011, supp}, the interacting part of the action is given by
\bea\label{interaction}
	S_{\text{int}} & = & \int_0^\infty dt \int_x \Big[ - \frac{u_c}{4!}2\phi_c^3\phi_q - \frac{u_q}{4!}2\phi_q^3\phi_c  \nn\\
 	&& - \frac{g_c}{\sqrt2} \phi_c \bar\Psi  \Psi -\frac{g_q}{\sqrt2} \phi_q \bar\Psi \tau^x \Psi \Big],
\eea
where $\phi_{c/q}$ ($\psi_{c/q}$) refer to the classical and quantum fields~\cite{kamenev2011, supp}, respectively. $\Psi = (\psi_c, \psi_q)^{T}$, and $\tau$ is Pauli matrix acting on the classical/quantum fields. Summation over flavors of Dirac fields is implicit. Note that there is no symmetry that relates the classical and quantum fields. Thus, while they have the same bare values, $g_{c/q}$, $u_{c/q}$ are in principle independent coupling constants at lower energy scales.

The fast modes within the momentum shell, $k \in [\Lambda e^{-l}, \Lambda]$, are integrated out to generate the RG equation~\cite{janssen1988, mitra2015}. Here $l>0$ denotes the flow parameter. The interacting part Eq.~(\ref{interaction}) generates RG processes that renormalize the slow modes, namely, $S_{\text{eff}} = S_0 + \delta S$,
\bea
	\delta S \!=\! \langle S_\text{int} \rangle_> + \frac{i}2 \langle S_\text{int}^2 \rangle_> - \frac{1}6 \langle S_\text{int}^3 \rangle_> \!-\! \frac{i}{24} \langle S_\text{int}^4 \rangle_>,
\eea
where $\langle A \rangle_> \!=\! \int D\phi_> D\bar\Psi_> D\Psi_> A e^{i S_0}$ denotes the functional integration over fast modes, and the calculation is done to one-loop order.

In terms of dimensionless coupling constants, $\bar \Omega_0 \equiv \Omega_0 \Lambda^{-1}$, $\bar g_{c/q}^2 \equiv K_d \Lambda^{d-3} (1+\bar \Omega_0^2)^{\pm 1/2} g_{c/q}^2$ , $\bar u_{c/q} \equiv K_d \Lambda^{d-3} (1+\bar \Omega_0^2)^{\pm 1/2} u_{c/q}$, where $K_d \equiv \frac{2\pi^{(d+1)/2}}{(2\pi)^d \Gamma(\frac{d+1}2)}$, the  RG equations~\cite{supp} are,
\bea
\label{g}	\frac{d \bar g_{c,q}^2}{dl} &=& \Big[ D_{c,q}(\bar \Omega_0) - \frac38 \bar g_c^2  - \Big( \frac{N}4 + \frac38 \Big) \bar g_c \bar g_q \Big]\bar g_{c,q}^2, \\
\label{u}	\frac{d\bar u_{c,q}}{dl} \!&=& \! \Big[ D_{c,q}(\bar \Omega_0) - \frac{N}2 \bar g_c \bar g_q - \frac38  \bar u_c\Big] \bar u_{c,q} + 3N \bar g_c \bar g_q \bar g_{c,q}^2, \\
\label{quench}	\frac{d\bar \Omega_0}{dl} &=& \bar \Omega_0,
\eea
where $D_{c,q}(\bar \Omega_0) = 3-d \pm \frac{\bar \Omega_0^2}{1+\bar \Omega_0^2}$ can be understood as the bare scaling dimensions of the coupling constants. For $\bar \Omega_0 = 0+ \mathcal O(\epsilon)$, the RG Eqs~(\ref{g}), and~(\ref{u}) reduce to the corresponding equilibrium ones, giving rise to the usual equilibrium chiral Ising universality class (ECIFP, see e.g., ~\cite{scherer2018}). Also, from Eq.~(\ref{quench}) we know the quench parameter is a relevant variable with an unstable value $\bar \Omega_0 = 0$ and a stable one $\bar \Omega_0 \rightarrow \infty$. As a result, there is a crossover, indicated by $D_c(0)=3-d$ and $D_c(\infty)=4-d$.

{\it Shallow quench.---} For a shallow quench, $\bar\Omega_0 \ll 1$, it is expected that the universal prethermal dynamics is controlled by the ECIFP. Accordingly, the usual chiral Ising critical exponents are applicable to describe the prethermal dynamics. However, an additional initial slip exponent $\theta$ must be included to describe the effects induced by the initial condition.

\begin{figure}
	\subfigure[]{
		\includegraphics[width=2.4cm]{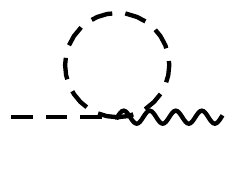}} \quad\quad
	\subfigure[]{
		\includegraphics[width=2.4cm]{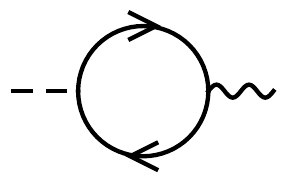}}
	\caption{\label{eta0}The Feynman diagrams that correct initial fields. The dashed line presents the retarded boson propagator, the wiggly line presents the quantum bosonic field, and the solid line indicates the fermion propagator.}
\end{figure}

To see this, we inspect the pre-quench action~\cite{supp},
\bea
	S_{s} =\frac{1}2 \int_k  \big( \omega_{0k} |\phi_{0q}({\bf k})|^2 - \frac{|\dot\phi_{0q}({\bf k})|^2}{\omega_{0k}}  \big),
\eea
where $\int_k \equiv \int \frac{d^dk}{(2\pi)^d}$, and $\phi_{0q}$ and $\dot \phi_{0q} \equiv \partial_t \phi_{0q}$ are initial fields defined at $t=0$. Because the sudden quench explicitly breaks time translation invariance, the Feynman diagrams shown in Fig.~\ref{eta0} lead to a correction of scaling of initial fields. The diagrams give a contribution of $\delta S_s = ( -\frac{\bar u_c}{16} + \frac{N}2 \bar g_c \bar g_q ) S_{s}$, shifting the anomalous dimension of the initial fields,
\bea
	\eta_{0}=- \frac{\bar u_c}{32} + \frac{N}4 \bar g_c \bar g_q.
\eea
The sign difference between these terms arises from the minus sign associated with the fermion loop in Fig.~\ref{eta0} (b).

This shift manifests in behavior of the retarded Green's function with initial fields $ i D_R({\bf k}, t, 0) = \langle \phi_c({\bf k},t) \phi_{q0}(-{\bf k})\rangle$. According to the scaling form of boson fields, we have $D_R({\bf k}, t, 0) = t^{1+ \theta} D_R({\bf k} t, 1, 0)$, where the critical initial slip exponent is $\theta \equiv -(\eta_b+\eta_0)$, and $\eta_b = \frac{\eta}2$ is the anomalous dimension of the boson field. For the retarded Green's function $D_R({\bf k}, t, t')$, when $t'$ is very close to zero, we can approximate $\phi_q(t) \approx \sigma(t) \phi_{0q}$~\cite{janssen1988}, with $\sigma(t)= b^{\eta_b - \eta_{0}} \sigma(bt) $, then
\bea
	D_R({\bf k}, t, t') = \frac{t^{1+ \theta}}{t'^{\eta+\theta}} \mathcal F({\bf k} t),
\eea
where $\mathcal F $ is a universal function.

The critical slip also manifests in the short-time scaling behavior of the order parameter. If the system is prepared with a finite order $M_0$, the order parameter shows early-time power-law time dependence set by the initial slip exponent. To set a finite order $\phi \sim M_0$, we modify the pre-quench Hamiltonian~\cite{supp}: 
\bea
H_{M_0} = \frac12 \int_x \Big[ \pi^2 + (\nabla \phi)^2 + \Omega_0^2 (\phi- M_0)^2 \Big].
\eea
Consequently, the pre-quench action is changed to
\bea
S_{s,M_0} = S_{s} + \int_x\frac{i M_0}{2\sqrt2}  \dot \phi_{0q},
\eea
and the order parameter in the presence of the initial magnetization is given by
\bea
	 M(t) &=& \langle \phi_{c}(t) e^{ -\int \frac{i M_0}{2\sqrt2} \dot \phi_{0q}} \rangle \nn\\
	 &=& \sum_{n=1}^\infty\int_{Y_n}  \frac{\big( \frac{-i M_0}{2\sqrt2} \big)^n}{n!}  \langle \phi_{c}(t) \prod_{i=1}^n \dot \phi_{0q}(y_i)\rangle,
\eea
where $\int_{Y_n}= \int_{y_1} ... \int_{y_n}$. Since the $n=0$ term vanishes,
\bea\label{magnetization}
	M(t) = M_0 t^{\theta} \mathcal M(t^{\mathcal D_{\Omega_0}+ \frac\eta2 +  \theta}M_0 ),
\eea
where $\mathcal D_{\Omega_0}$ is the canonical scaling dimension of $\phi_c$, and $\mathcal M$ is a universal function. Notice that Eq.~(\ref{magnetization}) validates after a non-universal initial time scale $\Lambda^{-1}$.

The critical exponents including the initial slip exponent of ECIFP are shown in Table~\ref{exponent}. It is interesting to note that, unlike the pure bosonic case, the initial slip exponent can be rendered negative if the number of fermion flavors $N$ is large enough (at one-loop, $N=1$ is already enough to make $\theta<0$), leading to a unique signature massless fermions in prethermal dynamics. The reason is that the large fluctuations of the fermion drive the critical point to be larger than its mean-field value.

{\it Deep quench.---}Next, for a deep quench, $\bar\Omega_0 \gg 1 $, the system is initially prepared far away from critical point and then suddenly quenched to the dynamical critical point. Unlike for a shallow quench, here $\epsilon = 4-d$. The critical and quantum coupling constants are significantly different already at the level of their bare couplings: $D_c(\infty) = 4-d$ while $D_q(\infty) = 2-d$. In $d=4-\epsilon$ dimensions, the quantum coupling constants are highly irrelevant, i.e., $g_q^*=u_q^*=0$. So, the low-energy dynamics are largely controlled by the classical coupling constants. From Eq.~(\ref{g}), the flow equations reduce to
\bea\label{non}
\frac{d \lambda}{dl}= \epsilon \lambda - \frac38 \lambda^2,
\eea
for $\lambda = \bar g_c^2, \bar u_c$.

Remarkably, Eq.~(\ref{non}) predicts a new fixed point with a $\bar g_c^2= \bar u_c=\frac{8 \epsilon}{3}$. We call this new fixed point the \textit{dynamical chiral Ising fixed point} (DCIFP) and observe that it is a \textit{non-thermal fixed point}. For a conventional fixed point describing thermal criticality, the Yukawa coupling is irrelevant and the fermionic degrees of freedom decouple at low energies because their Matsubara frequencies do not include zero. For example, at finite temperatures, the classical thermal phase transition for the Dirac system, whose quantum phase transition at the zero temperature is the chiral Ising universality class, belongs to the conventional Ising universality class~\cite{Stephanov,Hesselmann}. However, for the DCIFP, $g_c\neq 0$, indicating the participation of the fermions. It was argued that in the pure bosonic model the dynamical fixed point bears similarities to the equilibrium thermal one. Here, we find that the dynamical phase transition can be quite different from the thermal one, with the physics of the dynamical fixed point originating in the interplay between the quenched boson fields and the gapless fermion fields.

The critical exponents associated with the DCIFP are listed in Table~\ref{exponent}. The ECIFP Yukawa coupling does not vanish, and it leads to a nonvanishing fermionic scaling dimension $\eta_f$. This anomalous dimension is observable, for example, in the energy $\varepsilon$  dependence of fermion density of states, $n(\varepsilon) \propto \varepsilon^{1+2\eta_f} $, for time-scales less than the thermalization time $t \ll t_\text{th}'$. The thermalization time is given by~\cite{supp} $t'_\text{th} \sim \Lambda^{-1}\epsilon^{-2} $. We emphasize that this non-equilibrium universality class does not have a classical counterpart and only exists in non-equilibrium states of matter.

\begin{figure}
	\subfigure[]{
		\includegraphics[width=4.1cm]{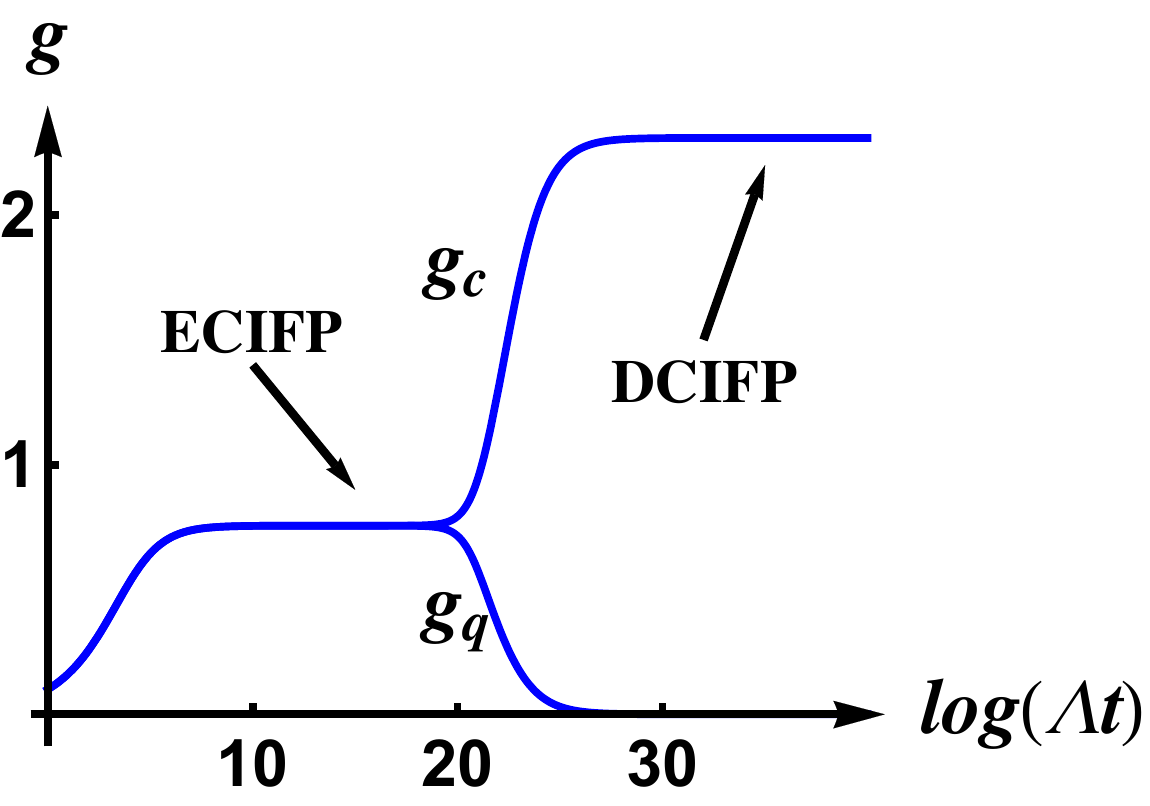}}
	\subfigure[]{
		\includegraphics[width=4.2cm]{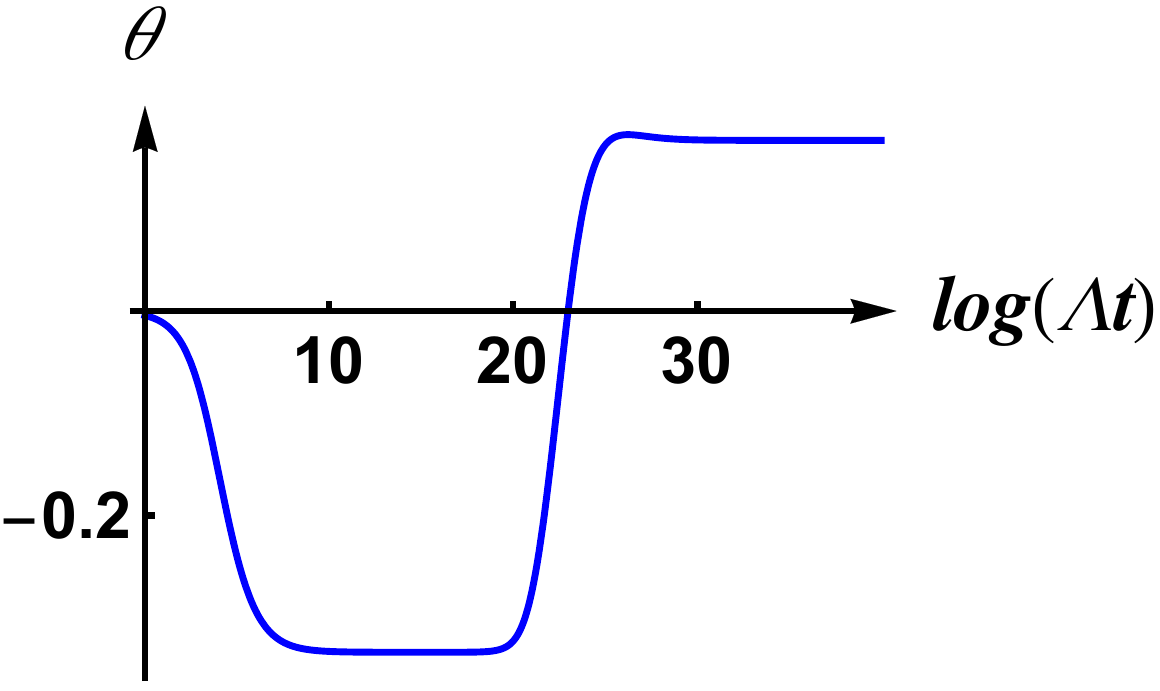}}
	\caption{\label{crossover} Temporal crossover from the near-equilibrium chiral Ising universality class to the non-equilibrium chiral Ising universality class. (a) The crossover of the quantum and classical Yukawa constants $g_{c/q}$. The two plateaus implies the two universality classes. (b) The crossover of initial slip exponent $\theta$. It should be noted that $\theta$ is rendered negative in the near-equilibrium chiral Ising universality class. In above plot, we choose $d=2, N=4$.}
\end{figure}

{\it Temporal crossover between two universality classes.}---Above we have shown that depending on the quench variable $\bar \Omega_0$, the universal prethermal dynamics is controlled by the ECIFP or the DCIFP. Since $\bar\Omega_0$ is a relevant quantity, the prethermal dynamics near the ECIFP resulting from a shallow quench is unstable at low energies, and the dynamics will flow to a scaling regime governed by the DCIFP.

Note that the crossover RG flow can manifest itself as a temporal crossover. Because $t\sim \mu^{-1}$ with $\mu$ being the energy scale in the question, the running RG parameter is effectively set by $l = \log \Lambda t$. In this sense, the RG flow can be directly seen in time-resolved experiments. Figure~\ref{crossover} shows the temporal crossover between the two scaling regimes. The two plateaus of the Yukawa coupling $g_{c/q}$ in Fig.~\ref{crossover} (a) indicate the RG flow near the ECIFP and the DCIFP, respectively. Since the bare values of $g_{c/q}$ are the same, the flows are identical in the scaling regime controlled by the ECIFP. However, at longer times, a non-zero $\bar \Omega_0$ drives a difference between $g_{c/q}$ through the bare scaling dimensions $D_{c/q}$. The quantum Yukawa coupling $g_q$ flows to zero while the classical $g_c$ flows to another finite value corresponding to the DCIFP. Intriguingly, Fig.~\ref{crossover} (b) shows the temporal crossover of the initial slip exponent $\theta$, where $\theta<0$ for large enough $N$ in the regime governed by the ECIFP and $\theta>0$ in the regime governed by the DCIFP.

{\it Discussion.---}The observability of these non-equilibrium universal scaling behaviors relies on a long thermalization time scale. In the present case, the thermalization can be understood as the generation of the dissipative term, $\gamma \phi_q \dot \phi_c$, in the RG flow~\cite{mitra2015,Gambassi2015,mitra2016,marino2017,mitra2011}. Accordingly, the RG equation of $\bar \gamma \equiv \gamma \Lambda^{-1}$ effectively sets the thermalization time scale. Initially, the prequench Hamiltonian describes a noninteracting system without dissipation, i.e., $\gamma=0$. The postquench interactions generate inelastic processes and render the dissipative coupling $\gamma$ nonzero.
For a shallow quench~\cite{supp}, the thermalization time is $t_\text{th} = \Lambda^{-1} \Big(1+ N \frac{1-\epsilon}{16\alpha_3 \epsilon^2} \Big)^{\frac{3+N}{3+(1-\epsilon)N}}$,
where $\alpha_3>0$, and $\epsilon = 3-d$. One finds that the thermalization time scale is extremely long at large $N$, even when $\epsilon = 1$. Indeed, the thermalization time scale is $t_\text{th} \propto (1+ \frac3{16 \alpha_3 })^N$ for $d = 2$ and $N \gg 1$. This feature is due to the integrability of zero-temperature Gross-Neveu-Yukawa model in the infinite-$N$ limit. On the other hand, for the deep quench, the thermalization time is $t_\text{th}' = \Lambda^{-1}\Big(1+ \frac9{64 \alpha_1 \epsilon^2}  \Big)$,
where $\alpha_1>0$ and $\epsilon = 4-d$. For a deep quench, the thermalization time scale is only controlled by $\epsilon$ because the boson cannot efficiently mediate an interaction between different fermion flavors at the DCIFP.

The results obtained above provide several sharp experimental signatures, in particular, the negative initial slip exponent for a shallow quench and its sign reversal in the temporal crossover to a deep quench arise from the fermion fields. Recently, the generation and detection of nonequilibrium dynamics has been demonstrated in various systems. In particular, in cold atom systems, dynamical scaling has been observed in experiments~\cite{Oberthaler2015,Eigen2018,Erne2018,Oberthaler2018}. Furthermore, the real-time dynamics has been measured in a tunable honeycomb lattice, which hosts gapless Dirac fermions~\cite{Greif2015,Bloch2017}. Accordingly, the prethermal dynamics studied here appear within experimental reach. It would also be interesting if some aspect of this physics survived to strong coupling, where it might serve as a signature of emergent fermions in candidate spin liquid materials.

To summarize, we studied the prethermal dynamics of Dirac systems. Two fixed points were found to govern the quench dynamics, the usual equilibrium chiral Ising fixed point and a new dynamical chiral Ising fixed point. The latter fixed point is non-thermal due to a non-trivial coupling between the boson and fermion degrees of freedom. The initial slip exponent was calculated for both shallow and deep quenches, and the negative initial slip exponent in the shallow case and its sigh change in the temporal crossover to the deep case provide a sharp signature of prethermal dynamics in Dirac systems.

{\it Acknowledgement:} We thank H. Y. Xie and H. Yao for helpful discussions. This work is supported in part by the NSFC under grant 11825404 (SKJ and SY), the Simons Foundation via the It From Qubit Collaboration (SKJ and BS), and the Department of Energy award number de-sc0017905 (BS). S.Y. is also supported in part by China Postdoctoral Science Foundation (Grant No. 2017M620035).

\begin{widetext}

\section{Supplemental Material}
\renewcommand{\theequation}{S\arabic{equation}}
\setcounter{equation}{0}
\renewcommand{\thefigure}{S\arabic{figure}}
\setcounter{figure}{0}
\renewcommand{\thetable}{S\arabic{table}}
\setcounter{table}{0}

\subsection{A. Propagators in Keldysh contour}

To prepare the initial state away from the transition point in the disordered phase, we use the Keldysh contour as indicated by Fig.~\ref{contour}. The vertical line in Fig.~\ref{contour} indicates the preparation of the pre-quench state.  The action is $e^{-S_s + i S_b}$, where
\bea
	S_s &=& \int_0^\beta d\tau L_E[\phi_i, \psi_i^\dag, \psi_i], \\
	S_b &=& \int_0^\infty dt \big( L[\phi_+, \psi_+^\dag, \psi_+ ] - L[\phi_-, \psi_-^\dag, \psi_- ] \big),
\eea
are the ``surface" and ``bulk" action. The vertical line is a constant time slice while horizontal lines stand for the bulk of spacetime. The Lagrangians are
\bea
\label{surfaceLagrangian}		L_E[\phi, \psi^\dag, \psi] &=& L_{E,b}[\phi] + L_{E,f}[\psi^\dag, \psi], \\
	L_{E,b} [\phi] &=& \frac12 \int_x  [(\partial_\tau \phi)^2 + (\nabla \phi)^2 + \Omega_0^2 \phi^2], \\
	L_{E,f}[\psi^\dag, \psi] &=& \int_x \psi^\dag (\partial_\tau + \mathcal H) \psi, \\
	L[\phi, \psi^\dag, \psi] &=& \int_x \big( \frac12 [\dot \phi^2 - (\nabla \phi)^2 - r \phi^2] + \psi^\dag (i \partial_t - \mathcal H) \psi + g \phi \psi^\dag \sigma^z \psi + \frac{u}{4!} \phi^4  \big),
\eea
where $\int_x \equiv \int d^d x$, $d$ is the spatial dimension, and $\dot \phi \equiv \partial_t \phi$. In this paper, we mainly focus on $d=2$ dimensions. $\phi$ is a real boson serving as an order parameter, and $\psi_\alpha$ ($\psi_\alpha^\dag$), $\alpha= 1, ..., N$, denotes the annihilation (creation) operator of two-component Dirac fermion for $\alpha$ flavor. The summation over $N$ flavors is assumed. The Dirac Hamiltonian is $\mathcal H = -i \sigma \cdot  \nabla$, where $\sigma \cdot  \nabla \equiv \sum_{i=1,2} \sigma_i \partial_i$ and $\sigma_i$ is the Pauli matrix. $\Omega_0^2$ and $r $denote the pre- and post-quench mass of the real boson, respectively. The quench from disordered phase towards the transition point obeys $\Omega_0^2 \gg r $. In the bulk Lagrangian, $g$ and $u$ denote the Yukawa coupling and four-boson coupling, respectively.

To get the propagators in the Keldysh contour, we perform a Keldysh rotation, i.e.,
\bea
\label{quantumBoson}	\phi_{c/q} &=& \frac{1}{\sqrt{2}} (\phi_+ \pm \phi_-), \\
\label{quantumFermion1} 	\psi_{c/q} &=& \frac{1}{\sqrt{2}} (\psi_+ \pm \psi_-), \\
\label{quantumFermion2}	\psi^\dag_{c/q} &=& \frac{1}{\sqrt{2}} (\psi_+^\dag \mp \psi_-^\dag).
\eea
In terms of classical/quantum fields, the bulk action is $S_b = S_0 + S_\text{int}$, where we have separated the bulk action into a non-interacting part $S_0$ and an interacting part $S_\text{int}$. They are given by
\bea
\label{bulkAction}	S_0 &=& \int_0^\infty dt ( L_{0,f}[\Psi^\dag, \Psi] + L_{0,b}[\phi_{c/q}]), \\
	L_{0,f}[\Psi^\dag, \Psi] &=&  \int_k \Psi^\dag \left( \ba{cccc} i \partial_t - \sigma \cdot k & 0 \\ 0 & i \partial_t  - \sigma \cdot k \ea \right) \Psi, \\
	L_{0,b}[\phi_{c/q}] &=& \int_x (\dot \phi_c \dot \phi_q - \nabla \phi_c \nabla \phi_q -r \phi_c\cdot \phi_q ) =\int_k [\partial_t (\phi_q \dot \phi_c) - \phi_q (\partial_t^2 \phi_c + \omega_k^2 \phi_c)] , \\
\label{interactingAction}	S_{\text{int}} &=& \int_0^\infty dt \int_x \Big[ - \frac{g_c}{\sqrt2} \phi_c \Psi^\dag \sigma^z \Psi -\frac{g_q}{\sqrt2} \phi_q \Psi^\dag \sigma^z \tau^x \Psi - \frac{u_c}{4! N}2\phi_c^3\phi_q - \frac{u_q}{4! N}2\phi_q^3\phi_c \Big],
\eea
where $\int_k \equiv \int \frac{d^d k}{(2\pi)^d}$, $k\cdot \sigma \equiv \sum_{i=1,2} k_i \sigma^i $, $\omega_k \equiv \sqrt{k^2 + r} $, and $\Psi = (\psi_c, \psi_q)^T$.
Since there is no symmetry that relates the quantum and classical fields, we allow their couplings to evolve independently under RG. These independent coupling constants are  $g_c, g_q$ and $u_c, u_q$. $\tau$ is a Pauli matrix acting on the classical/quantum space.

Now we are ready to derive the propagators in the presence of the surface action. Since in both surface and bulk noninteracting action, Eqs.~(\ref{surfaceLagrangian},~\ref{bulkAction}), the fermion and boson are decoupled, we can derive their propagators independently. In terms of the classical/quantum fields, the propagators are given by
\bea
	\hat G = -i \langle \Psi \Psi^\dag \rangle = \left( \ba{cccc} G_R & G_K \\ 0 & G_A \ea \right) , \\
	\hat D = -i \langle \Phi \Phi^\dag \rangle = \left( \ba{cccc} D_K & D_R \\ D_A & 0 \ea \right).
\eea
In the following, we will first get the fermion propagator and then the boson propagator.

\begin{figure}[t]
	\includegraphics[width=6cm]{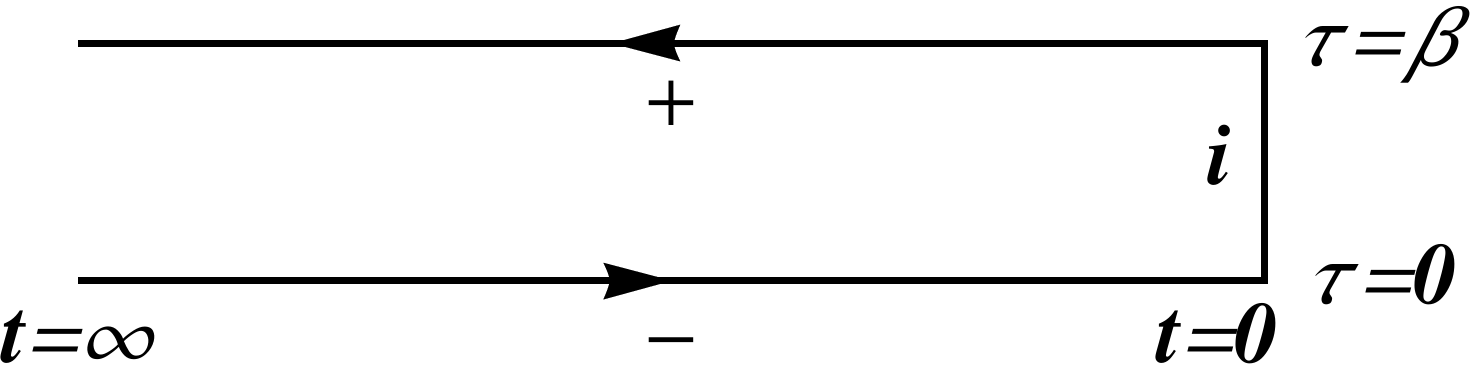}
	\caption{\label{contour}The Keldysh contour. The vertical line indicates the preparation of pre-quench state. $\pm$ denotes the fields evolving forwards/backwards along the horizontal real time axis, while $i$ implies the initial fields evolving in the vertical $t=0$ time slide. The boundary conditions are $\phi_-(0)= \phi_i(0)$, $\phi_+(0)=\phi_i(\beta)$, $\psi_-(0)=- \psi_i(0)$, and $\psi_+(0) = \psi_i(\beta)$. }
\end{figure}

We introduce source field $\xi_{c/q}$ coupling to $\psi_{c/q}$, and integrate out all dynamical fields to get the generating functional as a function of $\xi_{c/q}$. The generating functional is defined by
\bea
	W[\xi_{c/q}] = \ln \int D\mu_f \exp\Big[ -\int d\tau L_{E,f}[\psi^\dag_i, \psi_i] + i \int dt \big( L_{0,f}[\Psi^\dag, \Psi] + (\xi_c^\dag \psi_c + \xi_q^\dag \psi_q + H.c.) \big)\Big], \nn \\
\eea
where $D\mu_f \equiv D\psi^\dag_i D\psi_i D\psi^\dag_c D\psi_c D\psi^\dag_q D\psi_q$ is the functional measure of all dynamical fields. Integrating over $\psi^\dag_i$, we obtain the equation of motion of the initial fields
\bea\label{fermionEOM1}
	(\partial_\tau + \mathcal H) \psi_i = 0.
\eea
Since the equation of motion of fermion is a first order differential equation, only one boundary condition is needed. In order to simplify the notation, we define $\psi_0 \equiv \psi_i(0)$. It is straightforward to get the solution of Eq.~(\ref{fermionEOM1}), namely,
\bea
	\psi_i(\tau) = [P_+(k) e^{-k \tau} + P_-(k) e^{k \tau}] \psi_0,
\eea
where $P_\pm(k) \equiv \frac12 (1\pm \hat k \cdot \sigma)$, $\hat k \equiv (k_1/k, k_2/k)$, and $k = \sqrt{k_1^2 + k_2^2}$. Combining the boundary condition $\psi_-(0)=-\psi_i(0), \psi_+(0)= \psi_i(\beta)$ and Eqs.~(\ref{quantumFermion1},~\ref{quantumFermion2}), we have
\bea
\label{boundary1}	
\psi_c(0)= \frac1{\sqrt2}(P_+(k) e^{-k \beta} + P_-(k) e^{k \beta} -1 )\psi_0, \\
\label{boundary2}	
\psi_q(0)= \frac1{\sqrt2}(P_+(k) e^{-k \beta} + P_-(k) e^{k \beta} +1 )\psi_0, \\
\psi^\dag_c(0)= \frac1{\sqrt2} \psi_0^\dag (P_+(k) e^{-k \beta} + P_-(k) e^{k \beta} +1 ), \\
\label{boundary3}	
\psi^\dag_q(0)= \frac1{\sqrt2} \psi_0^\dag (P_+(k) e^{-k \beta} + P_-(k) e^{k \beta} -1 ).
\eea
After integrating the initial fields, we arrive at
\bea
	W[\xi_{c/q}] &=& \ln \int D\psi_0 D\psi_c^\dag D\psi_c D\psi_f^\dag D\psi_f \exp\Big[ i \int dt \big( L_{0,f}[\Psi^\dag, \Psi] + (\xi_c^\dag \psi_c + \xi_q^\dag \psi_q + H.c.) \big)\Big],
\eea
where we manipulate to get
\bea
	L_{0,f}[\Psi^\dag, \Psi] = \int_x \psi^\dag_c(i\partial_t -\mathcal H) \psi_c +i \partial_t (\psi_q^\dag \psi_q) -i (\partial_t \psi_q^\dag) \psi_q - \psi_q^\dag \mathcal H \psi_q.
\eea
Integrating over $\psi_c^\dag$ and $\psi_q$ leads to the equation of motion of $\psi_c$ and $\psi_q^\dag$ in the presence of source fields, namely,
\bea
\label{fermionEOM2}		(i\partial_t- \mathcal H  ) \psi_c +\xi_c =0, \\
\label{fermionEOM3}		- i \partial_t \psi_q^\dag - \psi_q^\dag \mathcal H +\xi_q^\dag = 0.
\eea
Recalling the boundary condition Eqs.~(\ref{boundary1},~\ref{boundary3}), it is straightforward to get solutions of Eqs.~(\ref{fermionEOM2},~\ref{fermionEOM3}),
\bea
	\psi_c(t) = \frac1{\sqrt2} [P_+(k) (e^{-k \beta}-1)  e^{-ikt}  + P_-(k) (e^{k \beta}-1)  e^{ikt}] \psi_0 - \int dt' G_R(k, t-t') \xi_c(t'), \\
	\psi_q(t)^\dag = \frac1{\sqrt2} \psi_0^\dag[P_+(k) (e^{-k \beta}-1)  e^{ikt}  + P_-(k) (e^{k \beta}-1)  e^{-ikt} ] - \int dt' \xi_q^\dag(t') G_A(k, t'-t),
\eea
where $G_{R/A}$ are retarded/advanced Green function given in the following,
\bea
	G_{R}(k,t-t') &=& -i\Theta(t-t') [e^{-i k (t-t')} P_+(k) + e^{i k (t-t')} P_-(k) ], \\
	G_{A}(k,t-t') &=& i \Theta(t'-t) [e^{-i k (t-t')} P_+(k) + e^{i k (t-t')} P_-(k) ].
\eea
Finally, integrating over $\psi_c$ and $\psi_q^\dag$ is amount to substitute the solution into the action,
\bea
e^{W[\xi_{c/q}]} &=& \exp-i \int dt dt' \int_k \big[ \xi_c^\dag(t) G_R(k, t-t') \xi_c(t') + \xi_q^\dag(t) G_A(k, t-t') \xi_q(t) \big] \\
	&&\times \int D\psi_0 \exp(\psi^\dag_{0q} \psi_{0q}) \exp \frac{i}{\sqrt2} \int dt \int_k \Big[ \xi_c^\dag [P_+ (e^{-k \beta}-1)  e^{-ikt}  + P_- (e^{k \beta}-1)  e^{ikt}] \psi_0   \nn \\
	&& + \psi_0^\dag[P_+ (e^{-k \beta}-1)  e^{ikt}  + P_- (e^{k \beta}-1)  e^{-ikt} ] \xi_q \big)\Big] ,
\eea
Given the boundary condition Eqs.~(\ref{boundary2},~\ref{boundary3}), we can integrate over $\psi_0$ to get the final result,
\bea
W[\xi_{c/q}] &=& \int dt dt' \int_k\Big( -i \big[ \xi_c^\dag(t) G_R(k, t-t') \xi_c(t') + \xi_q^\dag(t) G_A(k, t-t') \xi_q(t) \big] \\
 &&- \xi_c^\dag(t) \tanh \frac{\beta k}2 [ e^{-ik (t-t')} P_+ - e^{ik(t-t')} P_-] \xi_q(t').
\eea
from which we have the fermionic Keldysh propagator,
\bea
	G_K(k,t-t') &=& - i \tanh \frac{\beta k}{2} [e^{-i k(t-t')} P_+(k) - e^{i k(t-t')} P_-(k)].
\eea

Next, we will derive the bosonic propagator. We introduce a source field $j_{c/q}$ coupling to $\phi_{q/c}$, and integrate out all dynamical fields to get the generating functional,
\bea
	W[j_{c/q}] &=& \ln \int D\phi_i D\phi_c D\phi_q \exp\Big[- \int d\tau L_{E,b}[\phi_i] + i\int dt \big(L_{0,b}[\phi_{c/q}] + j_c \phi_q + j_q \phi_c \big) \Big].
\eea
To integrate out $\phi_i$, we first solve the equation of motion for $\phi_i$,
\bea\label{solution1}
	\phi_i(\tau) = \phi_-(0) (\cosh\omega_{0k} \tau - \coth\omega_{0k}\beta \sinh \omega_{0k} \tau ) + \phi_+(0) \frac{\sinh \omega_{0k}\tau}{\sinh \omega_{0k} \beta},
\eea
where $\omega_{0k}^2 \equiv k^2 + \Omega_0^2$ and we have used the boundary conditions $\phi_i(0)= \phi_-(0)$ and $\phi_i(\beta)=\phi_+(0)$. Then we plug the solution into the action to arrive at
\bea\label{preBoson}
	\int d\tau L_{E,b}[\phi_i] =\int_k \frac{\omega_{0k}}2 \big(\phi_{0c}^2 \tanh \frac{\beta \omega_{0k}}{2}+ \phi_{0q}^2 \coth \frac{\beta \omega_{0k}}{2} \big),
\eea
where $\phi_{0c} \equiv \frac1{\sqrt{2}} [\phi_+(0)+ \phi_-(0)]$ and $\phi_{0q}= \frac1{\sqrt{2}}[\phi_+(0) - \phi_-(0)]$. For later consideration, we take the imaginary time derivative of the solution Eq.~(\ref{solution1}), i.e.,
\bea
	\partial_\tau \phi_i(\tau) = \omega_{0k} \phi_-(0) (\sinh\omega_{0k} \tau - \coth\omega_{0k}\beta \cosh \omega_{0k} \tau ) + \omega_{0k}\phi_+(0) \frac{\cosh \omega_{0k}\tau}{\sinh \omega_{0k} \beta},
\eea
and using $\dot \phi_+(0) = i \partial_\tau \phi_i(\beta)$, $\dot \phi_-(0) = i \partial_\tau \phi_i(0)$, we can get
\bea\label{dotPhi}
	\dot \phi_{0c} = i \omega_{0k} \coth \frac{\beta\omega_{0k}}2 \phi_{0q},  \quad \dot \phi_{0q} = i \omega_{0k} \tanh \frac{\beta\omega_{0k}}2 \phi_{0c}.
\eea
As the equation of motion is second-order in derivatives, there are two independent fields, so Eq.~\ref{preBoson} can be equivalently expressed as
\bea\label{prePhiq}
	\int d\tau L_{E,b}[\phi_i] =\int_k \frac{\coth \frac{\beta \omega_{0k}}{2}}2 \big( \omega_{0k}\phi_{0q}^2 - \frac{\dot\phi_{0q}^2}{\omega_{0k}}  \big),
\eea

Integrating over $\phi_q$, we obtain the equation of motion in the presence of source field $j_c$,
\bea\label{bosonEOM}
	\partial_t^2 \phi_c + \omega_k^2 \phi_c = j_c(t).
\eea
It is straightforward to get the solution of Eq.~(\ref{bosonEOM}), i.e.,
\bea
	\phi_c(t) = \phi_{0c} \cos \omega_k t + \dot \phi_{0c} \frac{\sin \omega_k t}{\omega_k} + \int_0^t dt' \frac{\sin \omega_k (t-t')}{\omega_k} j_c(t').
\eea
We plug the solution into the action and integrate out $\phi_{0c}$ and $\phi_{0q}$ to get
\bea
	e^{W[j_{c/q}]}&=& \int D\phi_{0c} D\phi_{0q}   \exp\Big[ -\int_k \frac{\omega_{0k}}2 (\phi_{0c}^2 \tanh \frac{\beta \omega_{0k}}{2}+ \phi_{0q}^2 \coth \frac{\beta \omega_{0k}}{2})  - i \int_x \phi_{0q} \dot \phi_{0c} \\
	&& + i \int dt \int_x ( \phi_{0c}  j_q\cos \omega_k t + \dot \phi_{0c} j_q  \frac{\sin \omega_k t}{\omega_k}) + i \int dt dt' j_q(t) \Theta(t-t')\frac{\sin \omega_k (t-t')}{\omega_k} j_c(t')\Big] \nn \\
	&=& \exp\Big[ - \frac12 \int dt \int_k j_q(t) \frac{\coth\frac{\beta \omega_{0k}}{2}}{\omega_k} [K_+ \cos \omega_k (t-t')+ K_- \cos \omega_k(t+t')]j_q(t') \\
	&& + i \int dt dt' \int_k j_q(t) \Theta(t-t')\frac{\sin \omega_k (t-t')}{\omega_k} j_c(t') \Big],
\eea
where $K_\pm = \frac12 (\frac{\omega_k}{\omega_{0k}} \pm \frac{\omega_{0k}}{\omega_{k}} )$, and in the calculation we have used Eq.~(\ref{dotPhi}). From the generating functional, we can easily obtain the boson propagators,
\bea
	D_R(k,t-t') &=&  -\Theta(t-t')\frac{\sin \omega_k (t-t')}{\omega_k}, \\
	D_A(k, t-t') &=&  \Theta(t'-t)\frac{\sin \omega_k (t-t')}{\omega_k}, \\
	D_K(k,t,t') &=& -i\frac{\coth\frac{\beta \omega_{0k}}{2}}{\omega_k} [K_+ \cos \omega_k (t-t')+ K_- \cos \omega_k(t+t')],
\eea

In the following, we summarize both fermion and boson propagators derived above,
\bea
	G_{R}(k,t-t') &=& -i\Theta(t-t') [e^{-i k (t-t')} P_+(k) + e^{i k (t-t')} P_-(k) ], \\
	G_{A}(k,t-t') &=& i \Theta(t'-t) [e^{-i k (t-t')} P_+(k) + e^{i k (t-t')} P_-(k) ], \\
	G_K(k,t-t') &=& - i \tanh \frac{\beta k}2 [e^{-i k(t-t')} P_+(k) - e^{i k(t-t')} P_-(k)], \\
	D_R(k,t-t') &=&  -\Theta(t-t')\frac{\sin \omega_k (t-t')}{\omega_k}, \\
	D_A(k, t-t') &=&  \Theta(t'-t)\frac{\sin \omega_k (t-t')}{\omega_k}, \\
	D_K(k,t,t') &=& -i\frac{\coth\frac{\beta\omega_{0k}}2}{\omega_k} [K_+ \cos \omega_k (t-t')+ K_- \cos \omega_k(t+t')],
\eea
where $\omega_k \equiv \sqrt{k^2 + r}$, $\omega_{0k} \equiv \sqrt{k^2 + \Omega_0^2}$, and $K_\pm =  \frac12 (\frac{\omega_k}{\omega_{0k}} \pm \frac{\omega_{0k}}{\omega_{k}} )$. It is worth noting that the quench information is encoded in the bosonic Keldysh propagator, while all other propagators have time translation symmetry.

\subsection{B. Propagator in the zero temperature limit}
Since we are interested in the role of fermions in prethermalization dynamics near a critical point, we should focus on the zero-temperature limit. Since there is no symmetry to relate the classical and quantum fields, we allow the bosonic pre-quench masses to evolve independently. Namely, the pre-quench boson action Eq.~(\ref{preBoson}) can be generalized into
\bea
	\int d\tau L_{E,b}[\phi_i] =\int_k  \big(\frac{\omega_{0c}}2\phi_{0c}^2 + \frac{\omega_{0q}}2 \phi_{0q}^2  \big),
\eea
where $\omega_{0c} \equiv \sqrt{k^2 + \Omega_{0c}^2} $ and $\omega_{0q} \equiv \sqrt{k^2 + \Omega_{0q}^2}$. It will modifies the bosonic Keldysh propagator $D_K(k,t,t')$ by
\bea
K_\pm = \frac12 \Big(\frac{\omega_k}{\omega_{0c}} \pm \frac{\omega_{0q}}{\omega_{k}} \Big).
\eea
Because $\Omega_{0c}$ is the pre-quench a relevant scale, we take $1/\Omega_{0c} \rightarrow 0$ and define $\Omega_0 \equiv \Omega_{0q}$. As we mentioned, Under such simplification, $K_\pm = \pm \frac{\omega_{0k}}{2\omega_k}$. And the propagators at zero-temperature limit are given by
\bea
	G_{R}(k,t-t') &=& -i\Theta(t-t') [e^{-i k (t-t')} P_+(k) + e^{i k (t-t')} P_-(k) ], \\
	G_{A}(k,t-t') &=& i \Theta(t'-t) [e^{-i k (t-t')} P_+(k) + e^{i k (t-t')} P_-(k) ], \\
	G_K(k,t-t') &=& - i [e^{-i k(t-t')} P_+(k) - e^{i k(t-t')} P_-(k)], \\
	D_R(k,t-t') &=&  -\Theta(t-t')\frac{\sin \omega_k (t-t')}{\omega_k}, \\
	D_A(k, t-t') &=&  \Theta(t'-t)\frac{\sin \omega_k (t-t')}{\omega_k}, \\
	D_K(k,t,t') &=& -i\frac{\omega_{0k}}{2\omega_k^2} [\cos \omega_k (t-t')- \cos \omega_k(t+t')].
\eea

\subsection{C. Renormalization group calculation}

\begin{figure}
	\subfigure[]{\label{example1}
		\includegraphics[width=2cm]{bosonTwoPoint1}} \quad
	\subfigure[]{\label{example2}
		\includegraphics[width=2cm]{bosonTwoPoint2}} \quad
	\subfigure[]{
		\includegraphics[width=2.3cm]{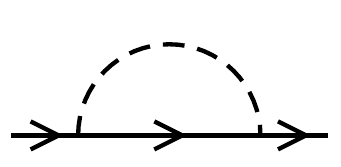}} \quad
	\subfigure[]{
		\includegraphics[width=2.3cm]{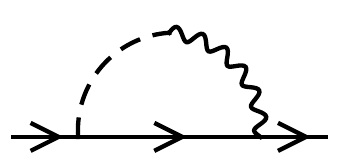}} \quad
	\subfigure[]{
		\includegraphics[width=2.3cm]{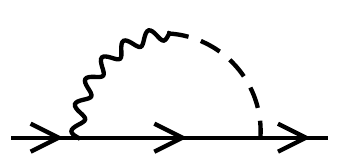}}
	\caption{\label{twoPoint}The Feynman diagrams that correct bosonic and fermionic two-point correlations.}
\end{figure}

The fast modes within the momentum shell, $[\Lambda e^{-l}, \Lambda]$, are integrated out to generate the RG equation. Here $\Lambda$ is the momentum cutoff and $l$ denotes the flow parameter which is a real number. To do that, we separate the fields into slow and fast modes, and the interacting part generates Feynman diagram that renormalize the slow modes, namely,
\bea
S_{\text{eff}} &=& S_0 + \delta S,\\
\delta S &=& \langle S_\text{int} \rangle_> + \frac{i}2 \langle S_\text{int}^2 \rangle_> - \frac{1}6 \langle S_\text{int}^3 \rangle_> - \frac{i}{24} \langle S_\text{int}^4 \rangle_> .
\eea
where $\langle A \rangle_> = \int D\phi_> D\Psi^\dag_> D\Psi_> A e^{i S_0}$ denotes the functional integration over fast modes. In the following calculation, we will keep the calculation up to one-loop order.

We are ready to calculate one-loop RG equations. Feynman diagrams shown in Fig.~\ref{twoPoint} lead to the renormalization, $\delta S_b^{(2)}$ and $\delta S_f^{(2)}$, of two-point functions of both fermionic and bosonic fields. To illustrate the calculation, we show in the following the calculation of first two Feynman diagrams in Fig.~\ref{twoPoint}:
\bea
\text{Fig.\ref{twoPoint}(a)}=  \int_0^\infty dt \int_x I_1(t) \phi_c \phi_q,
\eea
where
\bea
	I_1(t) &=& \frac{-u_c}4 \int_{d\Lambda} \frac{d^d k}{(2\pi)^d} i D_K(k,t,t) \nn \\
	&=& \frac{-u_c}4 K_d \int_{\Lambda e^{-l}}^\Lambda k^{d-1} dk \frac{\omega_{0k}}{2\omega_k^2} [1- \cos 2 \omega_k t] \nn \\
	&\approx& \frac{-u_c}4 K_d l \Lambda^d \frac{\omega_{0\Lambda}}{2\omega_\Lambda^2} [1- \cos 2 \omega_\Lambda t],
\eea
where $\omega_\Lambda \equiv \sqrt{\Lambda^2 + r}$, $\omega_{0\Lambda} \equiv \sqrt{\Lambda^2 + \Omega_0^2}$, $K_d \equiv \frac{A[S^{d-1}]}{(2\pi)^d}$, and $A[S^{d-1}]$ is the area of unit sphere $S^{d-1}$. And
\bea
\text{Fig.\ref{twoPoint}(b)}=  \int_0^\infty dt \int_p \phi_c I_2(p,t) \phi_q,
\eea
where
\bea
	I_2(p,t) &=& - \frac{i g_c g_q}{2} \int dt' \int_{d\Lambda} \Tr[i G(k,t,t') i G(k+p,t',t) \tau^x] \nn \\
	&\approx&  - \frac{i g_c g_q}{2} \int dt' \int_{d\Lambda} \Big[ \Tr[i G(k,t,t') i G(k,t',t) \tau^x] + \sum_j \Tr[i G(k,t,t') i \partial_{p_j}^2 G(k,t',t) \tau^x]  p_j^2 \Big] \nn \\
	&=&  - \frac{i g_c g_q}{2} K_d l \Lambda^d \Big[ \frac{2i}{\Lambda}(1-\cos 2\Lambda t)+ \frac{i[-1+(1+\frac23 \Lambda^2 t^2) \cos 2\Lambda t + \frac23 \Lambda t \sin 2\Lambda t]}{2\Lambda^3} p^2 \Big].
\eea
We will only keep time dependent contribution to $\int \phi_q \phi_c$ terms. Similarly, straightforward generalization of above calculations to other Feynman diagrams lead to the answer:
\bea
	\delta S_b^{(2)} &=&  \int_0^\infty dt  \Big( \int_x \phi_c \Big[ \frac{-u_c}4 K_d l \Lambda^d \frac{\omega_{0\Lambda}}{2\omega_\Lambda^2} (1- \cos 2\omega_\Lambda t)\Big] \phi_q \\
	&& +  \int_p \phi_c \Big[ g_c g_q K_d l \Lambda^d \frac{1-\cos2\Lambda t}\Lambda - \frac{g_c g_q}4 K_d l \Lambda^d \frac{p^2}{\Lambda^3}\Big]\phi_q \Big),
\eea
and
\bea
	\delta S_f^{(2)} = \int_0^\infty dt \int_p \Psi^\dag \Big[ - \frac{g_c^2}{12} K_d l \Lambda^d \frac{\omega_{0\Lambda}}{r^2}\Big( \frac{2(\omega_\Lambda-\Lambda)}{\Lambda} - \frac{r}{\omega_\Lambda^2} \Big)- \frac{g_c g_q}{12} K_d l \Lambda^d \frac{\omega_{\Lambda}}{r^2}\Big( \frac{2(\omega_\Lambda-\Lambda)}{\Lambda} - \frac{r}{\omega_\Lambda^2} \Big) \Big] p \cdot \sigma \Psi, \nn\\
\eea
\begin{figure}
	\subfigure[]{
		\includegraphics[width=2cm]{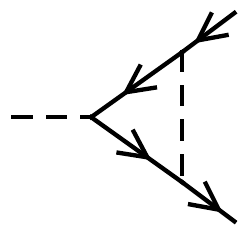}} \quad
	\subfigure[]{
		\includegraphics[width=2cm]{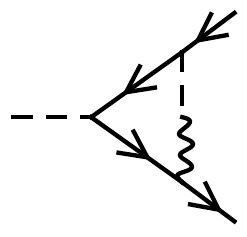}} \quad
	\subfigure[]{
		\includegraphics[width=2cm]{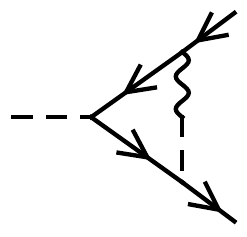}} \quad
	\subfigure[]{
		\includegraphics[width=2cm]{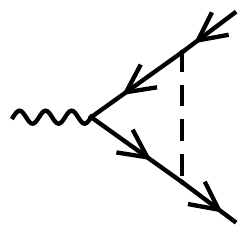}} \quad
	\subfigure[]{
		\includegraphics[width=2cm]{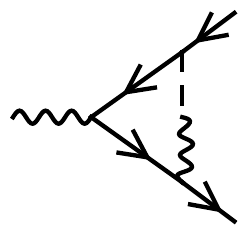}} \quad
	\subfigure[]{
		\includegraphics[width=2cm]{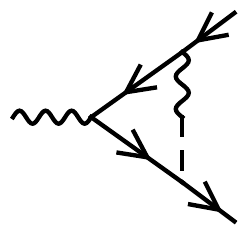}}
	\caption{\label{threePoint} The Feynman diagrams that correct three-point vertices.}
\end{figure}
Next, Feynman diagrams shown in Fig.~\ref{threePoint} lead to the renormalization, $\delta S^{(3)}$, of three-point vertices, namely,
\bea
	\delta S^{(3)} &=& \int_0^\infty dt \Big( \int_x \Big[\frac{g_c^3}{4\sqrt2} K_d l \Lambda^d \frac{\omega_{0\Lambda}}{r \omega_\Lambda} (\frac1{\Lambda}- \frac1{\omega_\Lambda}) + \frac{g_c^2 g_q}{2\sqrt2} K_d l \Lambda^d \frac{1}{r} (\frac1{\Lambda}- \frac1{\omega_\Lambda}) \Big] \phi_c \Psi^\dag \sigma^z \Psi \\
	&& + \int_x \Big[\frac{g_c^2 g_q}{4\sqrt2} K_d l \Lambda^d \frac{\omega_{0\Lambda}}{r \omega_\Lambda} (\frac1{\Lambda}- \frac1{\omega_\Lambda}) + \frac{g_c g_q^2}{2\sqrt2} K_d l \Lambda^d \frac{1}{r} (\frac1{\Lambda}- \frac1{\omega_\Lambda}) \Big] \phi_q \Psi^\dag \sigma^z\tau^x \Psi  \Big).
\eea

Finally, Feynman diagrams shown in Fig.~\ref{fourPoint} lead to the renormalization, $\delta S^{(4)}$, of four-point vertices, namely,

\bea
	\delta S^{(4)} =\int_0^\infty dt  \Big( \Big[\frac{u_c^2}{32} K_d l\Lambda^d \frac{\omega_{0\Lambda}}{\omega_\Lambda^4} - \frac{g_c^3 g_q}4 K_d l \Lambda^d \frac1{\Lambda^3} \Big] \phi_c^3 \phi_q +  \Big[ \frac{u_c u_q}{32} K_d l\Lambda^d \frac{\omega_{0\Lambda}}{\omega_\Lambda^4} - \frac{g_c g_q^3}4 K_d l \Lambda^d \frac1{\Lambda^3} \Big] \phi_q^3 \phi_c \Big), \nn \\
\eea

\begin{figure}
	\subfigure[]{
		\includegraphics[width=2.3cm]{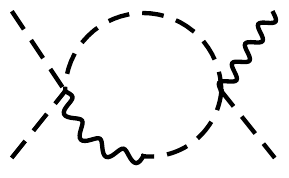}} \quad
	\subfigure[]{
		\includegraphics[width=2.3cm]{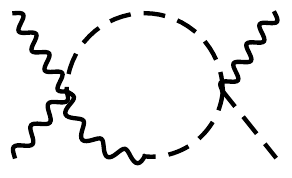}} \quad
	\subfigure[]{
		\includegraphics[width=2cm]{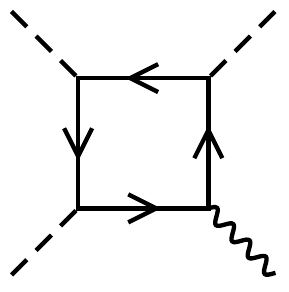}} \quad
	\subfigure[]{
		\includegraphics[width=2cm]{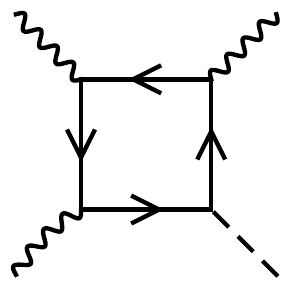}} \quad
	\caption{\label{fourPoint}The Feynman diagrams that correct four-point boson vertices.}
\end{figure}

Collecting the results from above calculations, the corrections from one-loop diagrams to the slow modes are given by,
\bea
	\delta S &=& \int_0^\infty dt \Big( \int_k \phi_c \Big[ \frac{-u_c}4 K_d l \Lambda^d \frac{\omega_{0\Lambda}}{2\omega_\Lambda^2} + g_c g_q K_d l \Lambda^d \frac1{\Lambda} - \frac{g_c g_q}4 K_d l \Lambda^d \frac{p^2}{\Lambda^3} \Big] \phi_q \nn \\
	&& + \int_k \Psi^\dag \Big[- \frac{g_c^2}{12} K_d l \Lambda^d \frac{\omega_{0\Lambda}}{r^2} \Big( \frac{2(\omega_\Lambda -\Lambda)}{\Lambda} - \frac{r}{\omega_\Lambda^2}\Big) - \frac{g_c g_q}{12} K_d l \Lambda^d \frac{\omega_{\Lambda}}{r^2} \Big( \frac{2 (\omega_\Lambda - \Lambda)}{\Lambda} - \frac{r}{\omega_\Lambda^2} \Big) \Big]  k\cdot \sigma \Psi \nn \\
	&& + \int_x \Big[ \frac{g_c^3}{4\sqrt2} K_d l \Lambda^d \frac{\omega_{0\Lambda}}{r\omega_\Lambda} \Big(\frac1{\Lambda}- \frac1{\omega_\Lambda} \Big) + \frac{g_c^2 g_q}{2\sqrt2} K_d l \Lambda^d \frac{1}{r} \Big(\frac1{\Lambda}- \frac1{\omega_\Lambda} \Big) \Big] \phi_c \Psi^\dag \sigma^z \Psi \nn \\
	&& + \int_x \Big[ \frac{g_c^2 g_q}{4\sqrt2} K_d l \Lambda^d \frac{\omega_{0\Lambda}}{r\omega_\Lambda} \Big(\frac1{\Lambda}- \frac1{\omega_\Lambda} \Big) + \frac{g_c g_q^2}{2\sqrt2} K_d l \Lambda^d \frac{1}{r} \Big(\frac1{\Lambda}- \frac1{\omega_\Lambda} \Big) \Big] \phi_c \Psi^\dag \sigma^z \tau^x \Psi \nn \\
	&& + \int_x \Big[ \frac{u_c^2}{32} K_d l \Lambda^d \frac{\omega_{0\Lambda}}{\omega_\Lambda^4} - \frac{g_c^3 g_q}4 K_d l \Lambda^d \frac1{\Lambda^3} \Big] \phi_c^3 \phi_q + \int_x \Big[ \frac{u_c u_q}{32} K_d l \Lambda^d \frac{\omega_{0\Lambda}}{\omega_\Lambda^4} - \frac{g_c g_q^3}4 K_d l \Lambda^d \frac1{\Lambda^3} \Big] \phi_c^3 \phi_q \Big). \nn \\
\eea
Now to get RG equations, we introduce the dimensionless coupling constants,
\bea
	&& \bar r \equiv r \Lambda^{-2}, \quad \bar \Omega_0 = \Omega_0 \Lambda^{-1}, \quad \bar g_c^2 \equiv K_d \Lambda^{d-3} (1+\bar \Omega_0^2)^{1/2} g_c^2 , \quad \bar g_q^2 \equiv  K_d \Lambda^{d-3} (1+\bar \Omega_0^2)^{-1/2} g_q^2, \\
	&& 	\bar u_c \equiv K_d \Lambda^{d-3} (1+\bar \Omega_0^2)^{1/2} u_c , \quad \bar u_q \equiv  K_d \Lambda^{d-3} (1+\bar \Omega_0^2)^{-1/2} u_q.
\eea
According to the renormalized action, the anomalous dimensions are given by
\bea
\eta_f = \frac{2(1+r)(\sqrt{1+r}-1)-r}{24r^2(1+r)} \bar g_c^2 + \frac{2(1+r)(\sqrt{1+r}-1)-r}{24r^2\sqrt{1+r}} \bar g_c \bar g_q, \quad \eta_b = \frac{N}8 \bar g_c \bar g_q.
\eea
And the RG equations are given by
\bea
	\frac{d\bar r}{dl} &=& 2\bar r- \frac{N}4 \bar g_c \bar g_q \bar r + \frac1{8(1+\bar r)} \bar u_c -\bar g_c \bar g_q, \\
	\frac{d\bar g_c^2}{dl} &=&  D_c(\bar \Omega_0)\bar g_c^2 - f_1(\bar r) \bar g_c^4 - f_2(\bar r) \bar g_c^3 \bar g_q, \\
	\frac{d\bar g_q^2}{dl} &=& D_q(\bar \Omega_0) \bar g_q^2 -f_1(\bar r) \bar g_c^2 \bar g_q^2 - f_2(\bar r)\bar g_c \bar g_q^3, \\
	\frac{du_c}{dl} &=& D_c(\bar \Omega) \bar u_c - \frac{N}2 \bar g_c \bar g_q \bar u_c - \frac38 \frac1{(1+\bar r)^2} \bar u_c^2 + 3N \bar g_c^3 \bar g_q, \\
	\frac{du_q}{dl} &=& D_q(\bar \Omega) \bar u_q - \frac{N}2 \bar g_c g_q \bar u_q - \frac38 \frac1{(1+r)^2} \bar u_c \bar u_q + 3N \bar g_c \bar g_q^3, \\
	\frac{d \bar \Omega_0}{dl} &=& \bar \Omega_0.
\eea
where $D_c(\bar \Omega_0) \equiv 3-d + \frac{\bar \Omega_0^2}{1+\bar \Omega_0^2}$, $D_q(\bar \Omega_0) \equiv 3-d - \frac{\bar \Omega_0^2}{1+\bar \Omega_0^2}$, and
\bea
	f_1(\bar r) = \frac{2(1+\bar r)(\sqrt{1+\bar r}-1)-\bar r}{6\bar r^2(1+\bar r)} + \frac{\sqrt{1+\bar r}-1}{2\bar r(1+\bar r)}, \\
	f_2(\bar r) = \frac{N}4 + \frac{2(1+\bar r)(\sqrt{1+\bar r}-1)-\bar r}{6\bar r^2\sqrt{1+\bar r}} + \frac{\sqrt{1+\bar r}-1}{2\bar r\sqrt{1+\bar r}}.
\eea
Notice that $\bar \Omega_0$ is relevant with an unstable value $\bar \Omega_0 = 0$ and a stable one $\bar \Omega_0 \rightarrow \infty$. To get some physical intuition, let's first assume the system is prepared close to critical point, namely, $\bar \Omega_0 = \bar r = 0+O(1/N)$. One expects the RG equations will reproduce the equilibrium ones. Indeed, the RG equations are given by
\bea
	\frac{d\bar g_c^2}{dl} &=&  (3-d) \bar g_c^2 -\frac38 \bar g_c^4 - (\frac{N}4 + \frac38)\bar g_c^3 \bar g_q, \\
	\frac{d\bar g_q^2}{dl} &=& (3-d) \bar g_q^2 -\frac38 \bar g_c^2 \bar g_q^2 - (\frac{N}4 + \frac38)\bar g_c \bar g_q^3 \\
	\frac{du_c}{dl} &=& (3-d) \bar u_c - \frac{N}2 \bar g_c \bar g_q \bar u_c - \frac38 \bar u_c^2 + 3N \bar g_c^3 \bar g_q, \\
	\frac{du_q}{dl} &=& (3-d) u_q - \frac{N}2 \bar g_c \bar g_q \bar u_q - \frac38 \bar u_c \bar u_q + 3N \bar g_c \bar g_q^3.
\eea
The similarity between $g_c$, $u_c$ and $g_q$, $u_q$ indicates only two of these coupling constants are needed, i.e.,
\bea
	\frac{d\bar g^2}{dl} &=&  (3-d) \bar g^2 -\frac38 \bar g^4 - (\frac{N}4 + \frac38)\bar g^4, \\
	\frac{du}{dl} &=& (3-d) \bar u - \frac{N}2 \bar g^2 \bar u - \frac38 \bar u^2 + 3N \bar g^4,
\eea
with a stable fixed point
\bea
	(g^*,u^*) = \Big(2 \sqrt{\frac{3-d}{3+N}}, \frac{4(3-d)\big(\sqrt{9+N(66+N)}+3-N \big)}{3(3+N)} \Big),
\eea
and critical exponents listed in Table~I. However, we should notice that $\bar \Omega_0$ is a relevant perturbation at this fixed point. From above lessons, it is straightforward to infer that the critical dynamics at short time after a soft quench  is controlled by the Gross-Neveu fixed point with a new nonequilibrium exponent---the critical initial slip.

On the other hand, when the system is initially prepared far away from critical point and suddenly quenched to the critical point, i.e., $\bar\Omega \rightarrow \infty$ and $\bar r = 0 + O(\epsilon)$, where $\epsilon = 4-d$. Now those critical and quantum coupling constants are very distinct from each other because the bare scalings, $D_c(\infty) = 4-d$ while $D_q(\infty) = 2-d$. The RG equations are given by
\bea
	\frac{d\bar g_c^2}{dl} &=&  (4-d) \bar g_c^2 -\frac38 \bar g_c^4 - (\frac{N}4 + \frac38)\bar g_c^3 \bar g_q, \\
	\frac{d\bar g_q^2}{dl} &=& (2-d) \bar g_q^2 -\frac38 \bar g_c^2 \bar g_q^2 - (\frac{N}4 + \frac38)\bar g_c \bar g_q^3, \\
	\frac{du_c}{dl} &=& (4-d) \bar u_c - \frac{N}2 \bar g_c \bar g_q \bar u_c - \frac38 \bar u_c^2 + 3N \bar g_c^3 \bar g_q, \\
	\frac{du_q}{dl} &=& (2-d) u_q - \frac{N}2 \bar g_c \bar g_q \bar u_q - \frac38 \bar u_c \bar u_q + 3N \bar g_c \bar g_q^3,
\eea
which possesses a stable fixed point given by
\bea
	(g_c^*, g_q^*, u_c^*, u_q^*) = \Big( 2 \sqrt{\frac{2 \epsilon}{3}}, 0, \frac{8 \epsilon}3,0  \Big).
\eea
The corresponding critical exponents are listed in Table~I.

\subsection{D. Critical initial slip exponent}

The surface action is given by
\bea
	S_{E,b}[\phi_i] =\int_k \frac{1}2 \big( \omega_{0k}\phi_{0q}^2 - \frac{\dot\phi_{0q}^2}{\omega_{0k}}  \big),
\eea
The time-dependent contribution will leads to renormalization of boundary field $\phi_{0f/0c}$. The time-dependent contribution is given by
\bea
	\delta S_s &=&  -i \int_x \int_0^\infty dt   \Big[ \frac{u_c}4 K_d l \Lambda^d \frac{\omega_{0\Lambda}}{2\omega_\Lambda^2} \cos 2\omega_\Lambda t - N g_c g_q K_d l \Lambda^d \frac{\cos2\Lambda t}\Lambda \Big] \phi_c \phi_q \\
	&\approx& -i \int_x (\dot \phi_{0c} \phi_{0q} + \phi_{0c} \dot \phi_{0q}) \int_0^\infty dt   t \Big[ \frac{u_c}4 K_d l \Lambda^d \frac{\omega_{0\Lambda}}{2\omega_\Lambda^2} \cos 2\omega_\Lambda t - N g_c g_q K_d l \Lambda^d \frac{\cos2\Lambda t}\Lambda \Big]  \\
	&=& \Big[ -\frac{u_c \omega_{0\Lambda} K_d l \Lambda^d}{16}  \frac{1}{\omega_\Lambda^4} + \frac{N}2 g_c g_q K_d l \Lambda^{d-3} \Big] \frac12 \int_k ( \omega_{0k} \phi^2_{0q} -  \frac1{\omega_{0k}}\dot \phi_{0q}^2),
\eea
which leads to anomalous dimension of initial fields, i.e., $\eta_{0}=- \frac{\bar u_c}{32(1+r)^2} + \frac{N}4 \bar g_c \bar g_q$. Let's consider the retarded Green's function with initial fields $ i D_R({\bf k}, t, 0) = \langle \phi_c({\bf k},t) \phi_{q0}(-{\bf k})\rangle$,
\bea
	D_R({\bf k}, t, 0) = b^{-1+\eta_b +\eta_0 } D_R({\bf k}/b, t b, 0).
\eea
Choosing $b=t^{-1}$, the retarded function has the scaling form
\bea
	D_R({\bf k}, t, 0) = t^{1+ \theta} D_R({\bf k} t, 1, 0).
\eea
Now we can introduce the cirtical initial slip $\theta \equiv -(\eta_b +\eta_0)$
To investigate the retarded Green's function $ i D_G({\bf k}, t, t')$ with $t'$ very close to zero, we can use $\phi_q(t) \approx \sigma(t) \phi_{0q}$, where $\sigma(t)= b^{\eta_b - \eta_{0}} \sigma(t b) $,
\bea
	D_R({\bf k}, t, t') =\sigma(t') D_R({\bf k}, t, 0) =b^{\eta_b - \eta_{0}} \sigma(t' b)  t^{1+ \theta} D_R({\bf k} t, 1, 0).
\eea
Setting $b= t'^{-1}$, we have
\bea
	D_R({\bf k}, t, t') = \frac{t^{1+ \theta}}{t'^{\eta+\theta}} \mathcal F({\bf k} t),
\eea
where $\mathcal F $ is a universal function.

A more experimental relevant quantity revealing the critical initial slip is the scaling form of order parameter. We change the boundary condition to set the initial order parameter to be $M_0$, namely, the boundary action is changed to
\bea
	S_{E,M_0} &=& \frac12 \int_x \int_0^\beta d\tau \Big[ (\partial_\tau \phi)^2 + (\nabla \phi)^2 + \Omega_0^2 (\phi- M_0)^2 \Big], \\
	&=& \int_x \frac12 \frac{M_0}{\sqrt2} \tanh \frac{\omega_{0k} \beta}2 \Big( \frac{\Omega_0^2}{\omega_{0k}} \phi_{0q} - \frac{\Omega_0^2}{\omega_{0k}} \phi_{0c} \Big) + S_{E, M_0=0}.
\eea
At zero temperature limit,
\bea
	S_{E,M_0} &=& \frac12 \int_x \frac{M_0}{\sqrt2} \Big( \frac{\Omega_{0q}^2}{\omega_{0q}} \phi_{0q} + i \frac{\Omega_{0c}^2}{\omega_{0c}^2} \dot \phi_{0c} \Big) + S_{E, b} \\
	&\approx & \frac12 \int_x\frac{M_0}{\sqrt2} \Big( \frac{\Omega_{0q}^2}{\omega_{0q}} \phi_{0q} + i \dot \phi_{0q} \Big) + S_{E, b} ,
\eea
where we have used Eq.~(\ref{dotPhi}) and take the limit $\Omega_{0c} \rightarrow \infty$ in the second line. It is easy to see the second term has a larger or equal scaling compared to the first term, so in order to get a scaling form, we can only keep the second term. The order parameter in the presence of the initial magnetization is given by
\bea
	 M(x, t) &=& \langle \phi_{c}(x, t) e^{ \int_y \frac{i M_0}{2\sqrt2} \dot \phi_{0q}} \rangle = \sum_{n=1}^\infty \int_{y_1,...,y_n} \frac{\big( \frac{i M_0}{2\sqrt2} \big)^n}{n!}  \langle \phi_{c}(x,t) \dot \phi_{0q}(y_1) ... \dot \phi_{0q}(y_n) \rangle.
\eea
It should be noticed that $n=0$ term vanishes. Then
\bea
	 M(x, t, M_0) &=&  \sum_{n=1}^\infty \int_{by_1,...,by_n}  b^{\mathcal D_{\Omega_0}+\eta_b} \frac{\big( \frac{i b^{\eta_0- \mathcal D_{\Omega_0}}M_0}{2\sqrt2} \big)^n}{n!}  \langle \phi_{c}(bx,bt) \dot \phi_{0q}(by_1) ... \dot \phi_{0q}(by_n) \rangle \\
	&=& M_0 b^{-\theta} M(b x, b t, b^{-\theta-\eta_b - \mathcal D_{\Omega_0}}M_0 ) =  M_0 t^{\theta} M(x/t, 1, t^{\mathcal D_{\Omega_0}+ \eta_b +  \theta}M_0 ), \nn
\eea
where $\mathcal D_{\Omega_0}$ is the bare scaling dimension of $\phi_c$, namely, $\mathcal D_0 = \frac{d-1}2$ and $\mathcal D_\infty = \frac{d-2}2$. Above equation leads to the result presented in the main text, $M(t, M_0) = M_0 t^{\theta} \mathcal M(t^{\mathcal D_{\Omega_0}+ \frac\eta2 +  \theta}M_0 ) $.

\subsection{E. Estimate of thermalization time}

\begin{figure}
	\subfigure[]{
		\includegraphics[width=2.3cm]{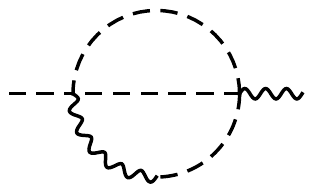}} \quad
	\subfigure[]{
		\includegraphics[width=2.3cm]{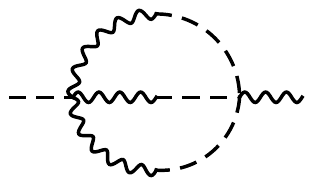}} \quad
	\subfigure[]{
		\includegraphics[width=2.3cm]{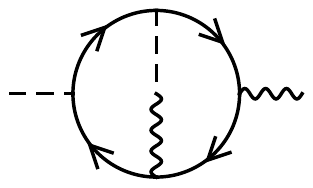}} \quad
	\subfigure[]{
		\includegraphics[width=2cm]{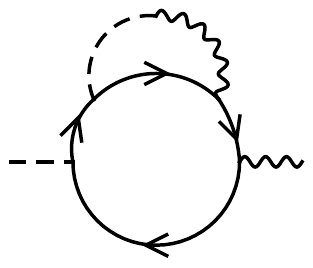}} \quad
	\subfigure[]{
		\includegraphics[width=2cm]{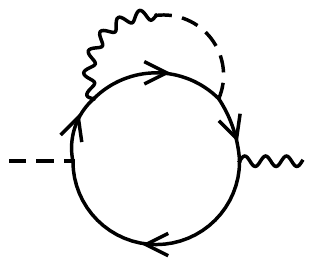}} \quad
	\caption{\label{twoLoop}The Feynman diagrams that lead to thermalization.}
\end{figure}

Thermalization is due to effectively irreversible processes. In an isolated system, although quantum time evolution is unitary, when we focus on one part of the system, say the low-energy modes in the Hilbert space, the other parts will effectively serve as a thermal bath. Then thermalization can be understood as the generation of a dissipative term, $\gamma \phi_q \dot \phi_c$, under the RG flow. The prequench Hamiltonian describes a noninteracting system without dissipation, so initially $\gamma=0$. When we quench the system, the interactions generate inelastic processes and drive the dissipative term $\gamma$ nonzero. Because the Keldysh boson propagator encodes the information of the quench protocol, the generation of a finite $\gamma$ is closely linked to the Keldysh boson propagator. The lowest order processes linking the dissipative term and Keldysh boson propagator come from the two-loop Feynman diagram shown in Fig.~\ref{twoLoop}, which gives an RG equation for $\bar \gamma \equiv \gamma \Lambda^{-1}$,
\bea
	\frac{d\bar \gamma}{dl} = (1-2\eta_b)\bar \gamma + \alpha_1 \bar u_c^2 + \alpha_2 \bar u_c \bar u_q - \alpha_3 N  \bar g_c^2 \bar g_q^2 ,
\eea
where $\alpha_i$ are some positive constants. To estimate the thermalization time, we plug the fixed point couplings into the above RG equation, and get the solution,
\bea
	\bar \gamma(l) = \frac{\alpha_3 N  \bar g_c^{2*} \bar g_q^{2*} - \alpha_1 \bar u_c^{\ast2} - \alpha_2 \bar u_c^* \bar u_q^*}{1-2\eta_b} (e^{(1-2\eta_b)l}-1).
\eea
Because $l=\log \Lambda t$, the thermal time scale is given by
\bea
	t_\text{th} = \Big( 1+ \frac{1-2\eta_b}{\alpha_3 N  \bar g_c^{2*} \bar g_q^{2*} - \alpha_1 \bar u_c^{\ast2} - \alpha_2 \bar u_c^* \bar u_q^*} \Big)^{1-2\eta_b}.
\eea
At the near-equilibrium chiral Ising fixed point, the thermalization time is
\bea
	t_\text{th} = \Lambda^{-1} \Big(1+ N \frac{1-\epsilon}{16\alpha_3 \epsilon^2} \Big)^{\frac{3+N}{3+(1-\epsilon)N}},
\eea
where $\epsilon = 3-d$. Apparently the thermalization time is extremely long at large $N$, even when $\epsilon = 1$. Indeed, the thermalization time scale is exponentially long $t_\text{th} \propto (1+ \frac3{16 \alpha_3 })^N$ for $\epsilon = 1 , N \gg 1$. On the other hand, at the dynamical chiral Ising fixed point, the thermalization time is
\bea
	t_\text{th}' = \Lambda^{-1}\Big(1+ \frac9{64 \alpha_1 \epsilon^2}  \Big),
\eea
where $\epsilon = 4-d$. Distinct from the near-equilibrium chiral Ising fixed point, thermalization time scale is only controlled by small $\epsilon$.

\end{widetext}

\end{document}